# Growth process and crystallographic properties of ammonia-induced vaterite

Qiaona Hu,[1,*] Jiaming Zhang,[1] Henry Teng,[2] and Udo Becker[1,*]

[1]Department of Geological Sciences, University of Michigan, 2534 C.C. Little, Ann Arbor, Michigan 48109, U.S.A.
[2]Department of Chemistry, George Washington University, 2121 I Street, Washington, D.C. 20052, U.S.A.

## Abstract

Metastable vaterite crystals were synthesized by increasing the pH and consequently the saturation states of $Ca^{2+}$- and $CO_3^{2-}$-containing solutions using an ammonia diffusion method. SEM and TEM analyses indicate that vaterite grains produced by this method are polycrystalline aggregates with a final morphology that has a sixfold-symmetry. The primary structure develops within an hour and is almost a spherical assemblage of nanoparticles (5–10 nm) with random orientation, followed by the formation of hexagonal platelets (1–2 μm), which are first composed of nanoparticles and that develop further into single crystals. As determined using transmission electron microscopy, these hexagonal crystallites are terminated by (001) surfaces and are bounded by {110} edges. The hexagonal crystals subsequently stack to form the "petals" (20 μm wide, 1 μm thick) of the final "flower-like" vaterite morphology. The large flakes gradually tilt toward the center as growth progresses so that their positions become more and more vertical, which eventually leads to a depression in the center. Since this sequence encompasses several morphologies observed in previous studies (spheres, hexagons, flowers etc.), they may actually represent different stages of growth rather than equilibrium morphologies for specific growth conditions.

## Introduction

Calcium carbonate ($CaCO_3$) occurs widely on the Earth's surface and represents the largest geochemical reservoir for carbon (Braissant et al. 2003; Wolf et al. 2007). There are three major polymorphs of $CaCO_3$: calcite, aragonite, and vaterite. Vaterite, the least thermodynamically stable polymorph of $CaCO_3$, has been of great interest because its crystallization is strongly associated with biogenic activities. For example, vaterite occurrences are widely documented in carbonate mineralization mediated by soil bacteria (Lian et al. 2006; Rodriguez-Navarro et al. 2007), in fresh-water cultured pearls from mussels (Addadi et al. 2006), as well as in inorganic tissues like gallstones and human heart valves (Kanakis et al. 2001; Lowenstam and Abbott 1975). In addition, vaterite can be an important precursor for other polymorphs, especially calcite, in biomineralization processes. In industrial settings, vaterite has been used extensively as a stabilizer in suspension polymerization, as material for regenerative medical applications, and as an encapsulating substance suitable for pharmaceutical applications (Fuchigami et al. 2009; Hadiko et al. 2005). Elucidating the vaterite growth mechanism and structure characteristics will expand our knowledge on how organisms control carbonate morphologies and polymorph selections and may improve the engineering applications of vaterite as an industrial material.

Various experimental conditions have been reported to promote vaterite synthesis. Facilitating factors include: supersaturation with respect to $CaCO_3$ (Colfen and Antonietti 1998; Grasby 2003; Kasuga et al. 2003; Kitamura 2001; Kralj et al. 1997; Ogino et al. 1987), the presence of certain organic additives and organic templates (Didymus et al. 1993; Fujiwara et al. 2010; Heywood et al. 1991; Hosoda et al. 2003; Jada and Verraes 2003; Kun Park et al. 2004; Lee et al. 2005; Malkaj et al. 2004; Rieke 1995; Rudloff and Colfen 2004; Shen et al. 2005) or inorganic additives (e.g., $NH_3$) (Gehrke et al. 2005; Liu and Yates 2006), and high pH (Spanos and Koutsoukos 1998). Different methods render different characteristics for vaterite crystallization, such as growth kinetics and crystal stability, among which the morphology may be the most uncertain characteristic because it displays a great deal of variety. The most intensively reported morphology is the spherical shape (Han et al. 2006a; Watanabe and Akashi 2008). Many others, such as fried-egg shape, flower-like shape, and hexagonal flake shape are also frequently observed (Fricke et al. 2006; Gehrke et al. 2005; Heywood et al. 1991). However, how these morphologies develop, what the texture of the resultant crystals is, and what the different shapes have in common, remains to be addressed. In this study, the crystallographic characteristics and growth process of a flower-like vaterite, induced by the ammonia-diffusion method, are investigated to shed light on the different stages of the crystallization mechanism of vaterite.

## Experimental methods

$CaCO_3$ crystals were cultivated using an ammonia diffusion method (Lian et al. 2006) in a closed container of 33 L at ambient conditions. Nine open plastic Petri dishes, each containing 45 mL of 0.002 $M$ $CaCl_2$-$NaHCO_3$ salt solutions, and one beaker, containing fresh $NH_4HCO_3$ powder, were placed in the container. The original pH of the salt solution was adjusted to 3.4 by adding 10% HCl such that initially, all Ca and carbonate species are in solution. Subsequent decomposition of $NH_4HCO_3$ in separate containers produced $NH_3$ gas, which diffused into the $CaCl_2$-$NaHCO_3$ salt solution, increasing the solution pH and, hence, the supersaturation with respect to $CaCO_3$. Two open Petri dishes containing 20% NaOH were placed in the glove box for absorbing the potentially existing $CO_2$, which would otherwise decrease the pH of the growth solution.

Preliminary tests show that the availability of ammonia strongly influences the polymorph composition of the precipitates. The optimal condition for a high percentage of vaterite crystallization was determined by running a series of parallel experiments of different ammonia diffusion rates by varying the amount and exposure area of the $NH_4HCO_3$ powder. It was found that nearly 100% vaterite yield (confirmed by both optical microscopy and using a Scintag X1 X-ray diffractometer) can be achieved using two 100 mL beakers each containing 25 g fresh $NH_4HCO_3$ powder. The pH of the salt solution increases from the original value, 3.4, to the final one, 9.6, within an hour. The supersaturations of the solutions calculated by MINTEQ at the final pH with respect to all three polymorphs, calcite, aragonite, and vaterite were 4.27, 3.95, and 2.97, respectively (MINTEQ version 2.51, Allison et al. 1991). The supersaturation state β is defined as

$$\beta = \ln\left(\frac{a(Ca^{2+})a(CO_3^{2-})}{K_{sp}}\right) \quad (1)$$

where $a(Ca^{2+})$ and $a(CO_3^{2-})$ are the activities of $Ca^{2+}$ and $CO_3^{2-}$, respectively. The final supersaturation state may fluctuate to some extent, but its influence on the growth procedure is negligible based on our observations in the repeating studies. Control experiments were conducted to examine the role of $NH_3$ on vaterite formation. This was done by adjusting the solution pH by adding NaOH instead of diffusing ammonia while maintaining the same rate of pH increase over the course of the experiment. It was found that calcite was the dominant phase in the control runs, indicating the ability of ammonia to promote vaterite formation.

Harvested vaterite crystals from each experimental run were dried and gold coated, and subsequently analyzed using field-emission-gun scanning electron microscopy (SEM, Hitachi S3200N). Specifically, crystals were collected every 30 min in the first four hours after the salt solution reached the final pH and every two hours thereafter until the crystals matured (in 16 h). The development of the vaterite morphology was also observed using in situ optical microscopy every two hours throughout the growth process.

The crystallographic properties of vaterite in the initial (within first hour) and final (after 16 h) stages were studied using transmission electron microscopy (TEM). Fine powder samples of vaterite were placed onto a holey carbon grid and analyzed using a JEOL 2010F instrument operating at 200 kV with minimized electron current density to avoid ionization. The chemical composition ($CaCO_3$) was confirmed by using energy-dispersive spectroscopy (EDS). Bright- and dark-field imaging was used to study the morphology in detail. The diffraction patterns and high-resolution TEM images were used to examine the crystal structures.

## RESULTS

### Effect of $NH_4^+/NH_3$ on vaterite formation

While calcite was the dominant phase in the control runs where NaOH was used to adjust pH, vaterite was nearly the only polymorphic phase in the ammonia diffusion experiments (Fig. 1). The $NH_3$ diffusion method is typically used to prepare calcite crystals; however, our work shows that the existence of $NH_3$ will greatly influence the initial polymorph composition of the precipitates. High percentages of calcite are only formed at low $NH_3$ diffusion rates, whereas vaterite becomes the major component when $NH_3$ diffuses fast and reaches certain concentrations (higher than 0.02 mol/L in this study) in the growth solution before $CaCO_3$ matures. More details on this topic are discussed in Hu et al. (in review).

### Evolution of crystal morphologies over time

SEM images (Fig. 2) of vaterite morphologies at different growth stages illustrate a dramatic change of shape over time. Spherical crystals about 2 μm in diameter were observed after the first hour (Fig. 2a), followed by the development of thin horizontal slabs branching out from the spheres within the next 30 min, transforming the original grain into a roughly hexagonal plate about 10 μm in diameter (Fig. 2b). Subsequent continuous growth of the slabs, with those at the peripheral area growing faster than the ones in the central region of the plate, caused the newly emerged layers to tilt progressively toward the center, resulting in a three-dimensional (3D) morphology with a pit in the center (Fig. 2d). The grain reached the size of about 20 μm in diameter after 4 h, and about 60 μm after 16 h when it fully matured. At the final stage, the vaterite grain displayed flaky crystal layers with a sixfold symmetry (Fig. 2e). The final leaves that formed are nearly vertical and surround a depression in the center (Fig. 2e, shown with higher magnification in Fig. 2f).

### Evolution from growth unit assemblage to final morphology

For the spherical vaterite grains grown in the first hour, bright- and dark-field images taken of the same area illustrate that the crystals are composed of randomly oriented nanoparticles that show no clearly defined shape (Figs. 3a and 3b). The corresponding high-resolution TEM image (Fig. 3c) reveals the discontinuity among the lattices of individual nanocrystals (5 nm). The fast Fourier transformation pattern of the high-resolution TEM image demonstrates a polycrystalline texture and confirms the crystal structure as vaterite.

In contrast, the more mature grains collected at later stages possess a more distinctive texture. Low-magnification SEM images show that the vaterite grains are composed of large flakes, each about 10–20 μm in diameter (Fig. 4b). High-magnification SEM and TEM images reveal that individual large flakes are formed by numerous small hexagonal platelets, each about 1 μm in diameter (Figs. 4c, 4d, and 5d). The corresponding diffraction patterns of the high-magnification TEM image demonstrate that the small hexagonal platelets are single crystals of vaterite with basal (001) faces and hexagonal boundaries parallel to {110}. A periodic arrangement of atoms of the hexagonal crystallites is revealed by the high-resolution TEM image shown in Figure 5c, displaying long-range order of atoms on the vaterite (001) surface.

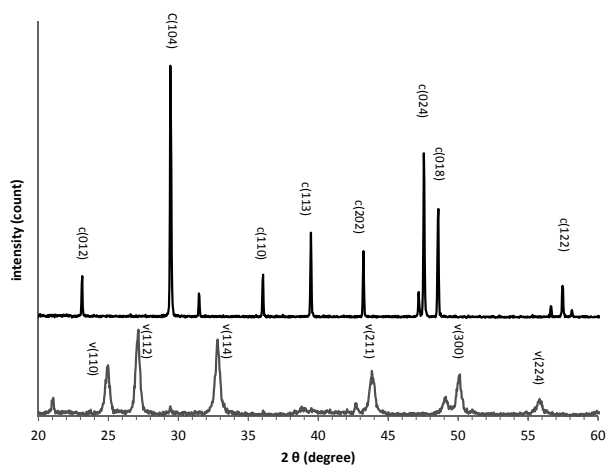

**FIGURE 1.** XRD results confirm that the precipitates are nearly pure vaterite in the experimental runs that applied the $NH_3$ diffusion method of high-evaporation rate of $NH_3$ (**top**). In contrast, calcite is the dominant phase in the control runs in the absence of $NH_3$ but by the application of dilute NaOH to increase pH (**bottom**). Numbers in parentheses indicate the Miller indices, and c and v denote calcite and vaterite, respectively.

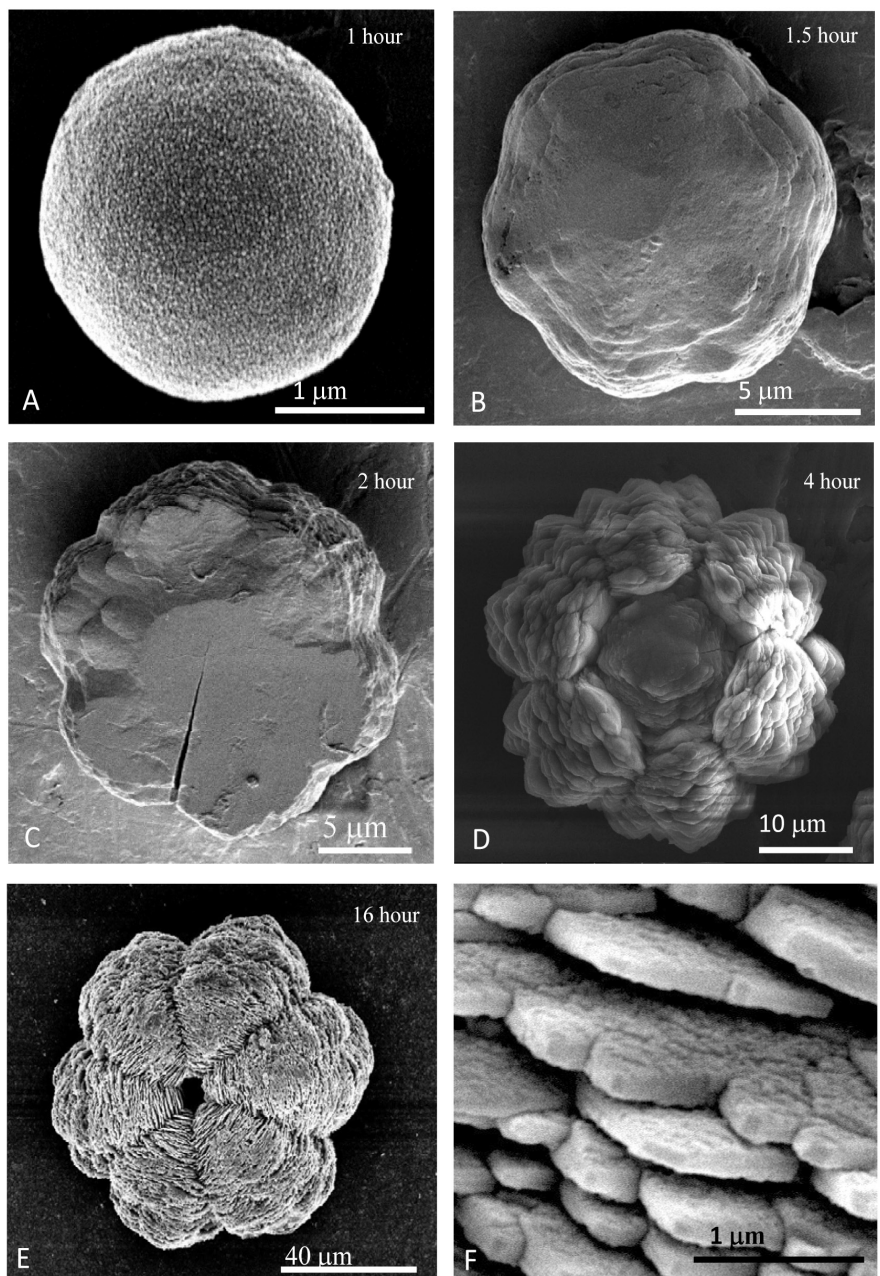

**FIGURE 2.** SEM images displaying the crystallization of vaterite grain as a function of time. The labels on the right corner of each image indicate the number of hours after which the crystals were harvested, as measured from the time the salt solution reached its final pH.

Although the individual hexagonal platelets in matured vaterite are single crystals, they appear as groups of nanoparticles of about 5–10 nm in diameter (Figs. 2f and 3). The size of these nanoparticles is similar to the particle sizes of the spherical grains formed within the first hour. Figure 6 shows a series of SEM images of the growth sequence of a vaterite grain, including details of its nanoparticulate structure. A series of high-magnification SEM images (Fig. 7) corresponding to the images in Figure 2 suggest how the vaterite crystals grew by means of reorganization of the clustering of nanoparticles. In the first hour, the vaterite grain was a spherical ball composed of randomly positioned/ oriented nanoparticles. Figure 7 indicates that in the next three hours, the nanoparticles began to align themselves to form hexagonal pieces, but the hexagonal shape is not well developed as the boundaries are jagged, and only show blunt angles around 120° between two sides. After four hours, the hexagonal shapes became increasingly clear, and the edges appeared more and more straight. In the final stage, these hexagonal pieces were well developed, with perfectly straight boundaries and the nanoparticles that were the original building blocks are tightly packed. The roughly spherical cores of the vaterite crystals preserved the disordered array of nanoparticles (Fig. 7), implying a lack of dissolution and re-crystallization after initial nucleation.

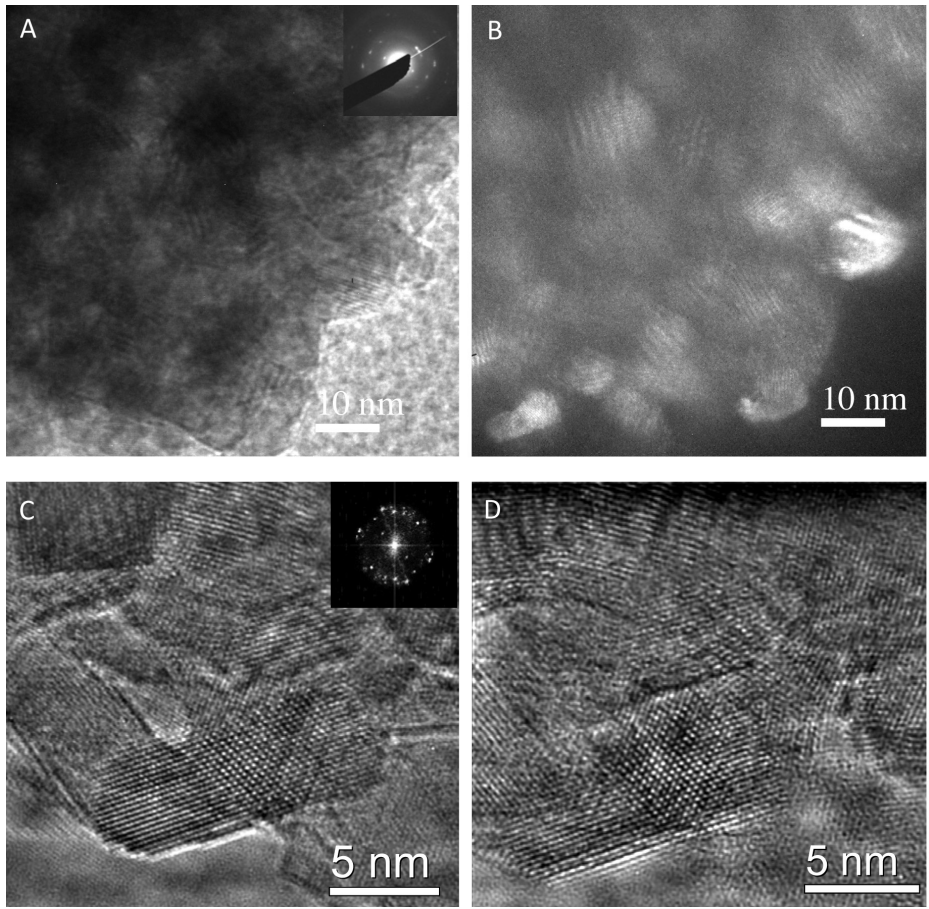

**Figure 3.** Bright- (**a**) and dark-field (**b**) TEM images, showing the vaterite nanocrystalline structure and lack of uniform orientations that formed within the first hour after the solution pH reached the maximum value. (**c**) Corresponding high-resolution TEM image. (**d**) Corresponding high-resolution TEM image after the 10 min electron beam bombardment, displaying a high resistance of nanocrystalline vaterite to the ionizing irradiation of the electron beam.

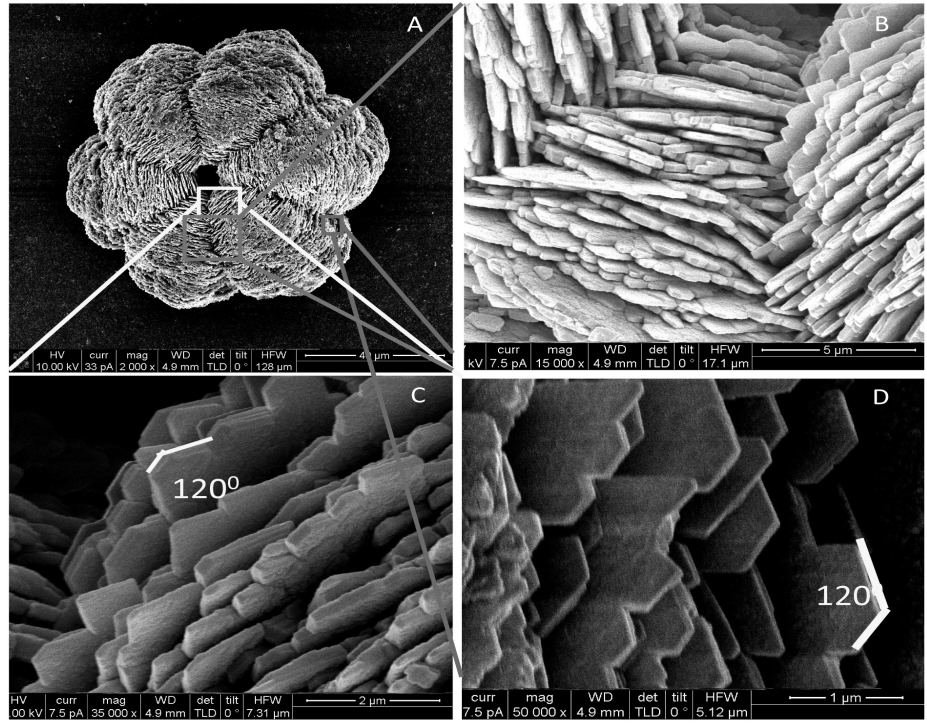

**Figure 4.** (**a** and **b**) Vaterite grain composed of large flakes (10–20 μm). (**c** and **d**) Each large flake is composed of numerous small hexagonal pieces (1 μm).

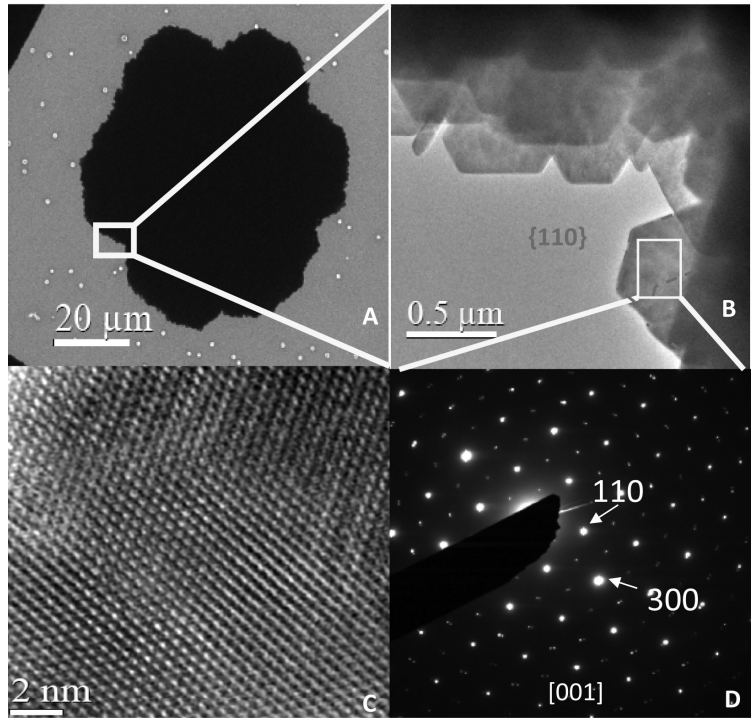

**Figure 5.** (**a** and **b**) Low- and high-magnification TEM images show the layered structure of vaterite and the unit bricks that are small hexagonal pieces, 1 μm. (**c**) A high-resolution TEM image shows atomic lattices of vaterite along [001], revealing an ordered arrangement of atoms. (**d**) Corresponding diffraction pattern along vaterite [001], demonstrating the hexagonal pieces shown in **b** terminates at {110}.

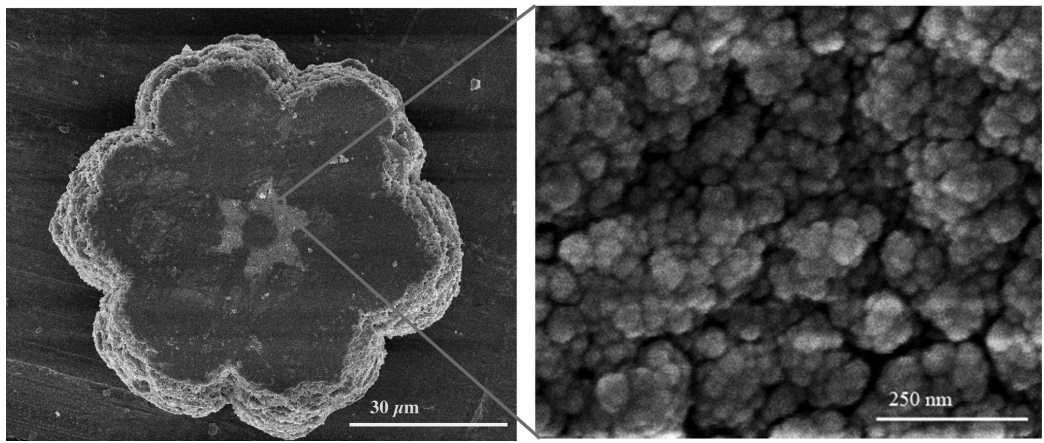

**Figure 6.** SEM image of a grain of an upside-down vaterite (**left**), showing the flat faces attaching to the Petri dish. In the center is the core area formed in the first two hours. The high-magnification SEM image of the core (**right**) shows the irregular texture and the aggregation of nanoparticles with random orientations.

An interesting phenomenon is that aggregates of nanoparticles formed in the early stage show an unusually high resistance to the ionizing irradiation of electron beam from TEM. $CaCO_3$, in general, is vulnerable to high temperatures and electron bombardment. In this study, after an irradiation at a rate of $1.25–2.5 \cdot 10^{12}$ electrons/cm$^2$, both single crystals of calcite (20–30 μm), or grown vaterite layers of crystalline structure (1–2 μm) begin to decompose. However, the nanocrystalline vaterite remained stable even after an irradiation at a rate of $1.12 \cdot 10^{14}$ electrons/cm$^2$. The high-resolution TEM image (Fig. 3d) displaying the crystallinity of nanoparticulate vaterite demonstrates that it is nearly intact in comparison to the original one before the $1.12 \cdot 10^{14}$ electrons/cm$^2$ irradiation (Fig. 3c).

It is interesting to note that in the TEM diffraction pattern (Fig. 5d), starting from the third nearest circle [i.e., (400) spots] to the transmitted spot (in the center), two separate spots can be detected at each location of the diffraction spot. The double-spot phenomenon is not clearly resolved in the first two circles close to the transmitted spot due to the high intensities of the lower-order diffraction spots. The distances between the two split spots increase with distance from the center (as illustrated). The contrasted color of the high-resolution TEM image (Fig. 5b) demonstrates that the area where the diffraction pattern was taken (as illustrated) locates two small hexagonal pieces. Therefore, the TEM diffraction pattern actually depicts the crystallographic structures of two crystallites, which illustrate two phenomena: (1) the two hexagonal platelets have the same structure, since the two sets of patterns are identical, and (2) a small degree of offset occurs between the two crystallographic orientations because one diffraction pattern is rotated slightly with respect to the other.

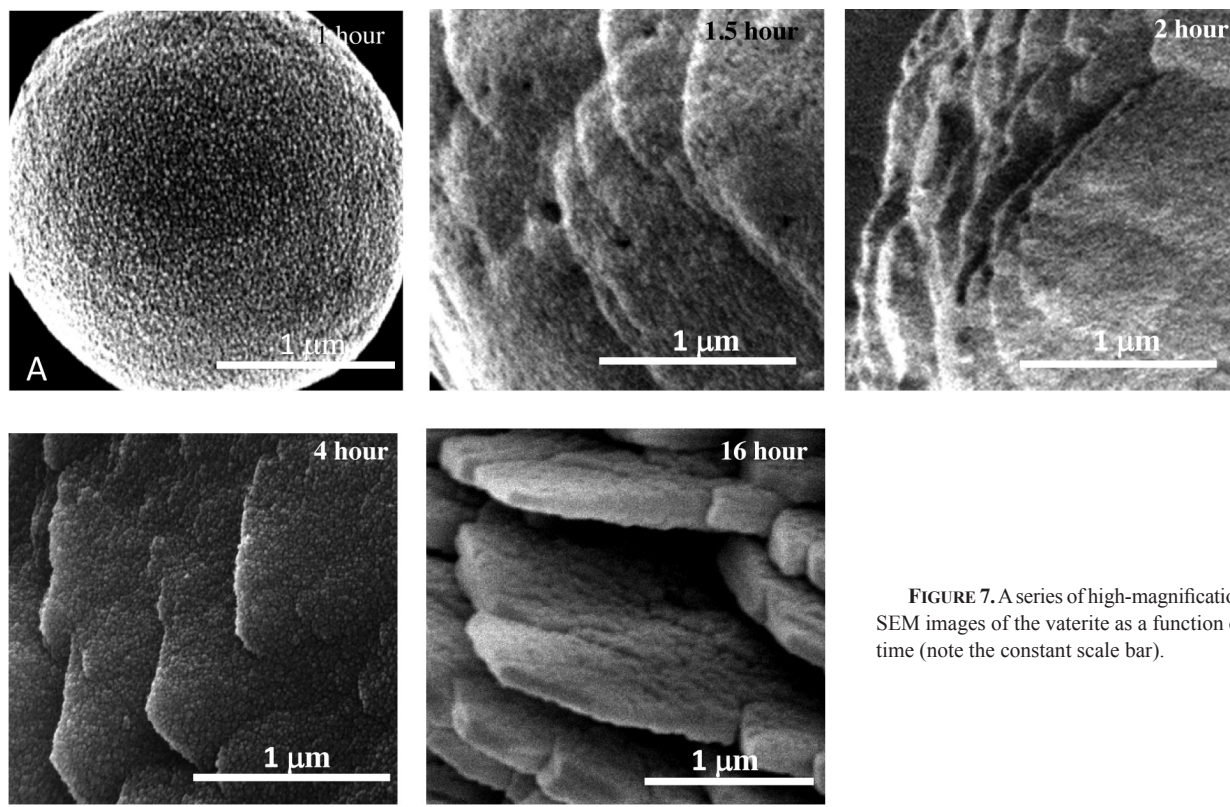

**Figure 7.** A series of high-magnification SEM images of the vaterite as a function of time (note the constant scale bar).

## Discussion

### The hierarchical structure and the development of vaterite grains

Based on the experimental observations, we propose a hierarchy-structure model of four levels for the texture of grown vaterite in this study. Individual grains of matured vaterite have a 3D structure of sixfold symmetry, like a flower with six petals (Figs. 2e and 2f). The primary structure is the basic level of this hierarchy and is the assemblage of nanoparticles (5–10 nm) (Fig. 5). The secondary structure is the next "level up" from the primary structure and is the stacking of hexagonal platelets (1–2 μm), which are initially composed of nanoparticles. These hexagonal pieces are single crystals. They grew together epitaxially and hence possess similar orientations. The tertiary structure is the arrangement of large stacked platelets (20 μm wide, 1 μm thick), which are built up by small hexagonal pieces, thus forming the six "petals" of the flower-like grain (Figs. 4c and 4d). The large flakes gradually tilt toward the center as growth progresses until their positions become vertical, eventually leading to the formation of a depression in the center. This quaternary structure is the last level of vaterite growth, referring to the sixfold arrangement of six petals (Figs. 4a and 4b).

### The building blocks of vaterite grains: nanoparticles

Given the structure characteristics of vaterite in different growth stages revealed by SEM and TEM analysis, we hypothesize that the nanoparticles (5–10 nm) are the building blocks of vaterite grains and the manner in which nanoparticles stack determines the overall shape of the vaterite.

Kamhi (1963) and Wang and Becker (2009) reported that the planar carbonate groups in vaterite may not have fixed positions but tend to randomly distribute over three orientations (the vectors normal to the carbonate planes are perpendicular to the $c$-axis and there is a 120° angle between these normal vectors). The uncertainty of carbonate-group orientation leads to inconsistent arrangements of atoms over the short range, which can easily result in other defects and structure mismatches, and hence interrupt the continuous growth of well-ordered larger single crystals of vaterite. This may explain why the development of vaterite is an assemblage of nanoparticles.

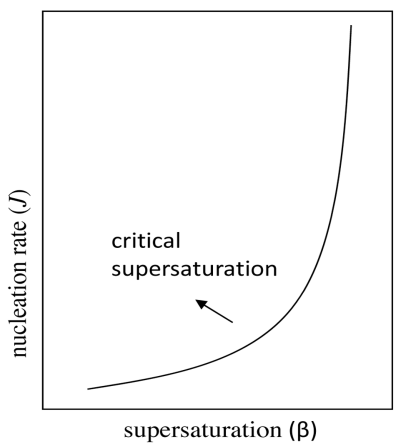

**Figure 8.** The nucleation rate increases dramatically after exceeding the critical supersaturation.

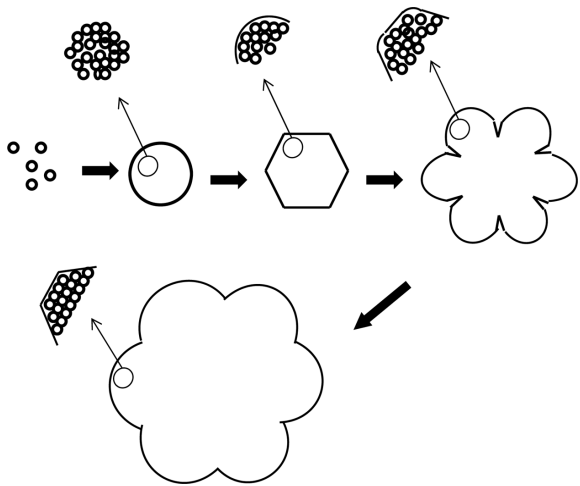

**Figure 9.** Schematic demonstrating the development of the sixfold flower-like vaterite grain. The 2D morphology of vaterite grain changes from spherical to hexagonal shape. The arrangement of the building blocks (nanoparticles) becomes increasingly ordered. See text for details.

Nanoparticles aggregate randomly in the early stage but subsequently line up along specific crystallographic orientations, resulting in a phase change from poorly crystalline to crystalline. In the first hour, clusters compact with random orientations, forming a sphere about a few micrometers in diameter. This may be due to the supersaturation that results in a high-driving force for $CaCO_3$ nucleation (Fig. 8). This phase with a random aggregation of nanoparticles, similar to an amorphous phase, is not the most energetically favorable one due to interfacial strain. Therefore, as the crystals precipitate, the supersaturation status decreases and the subsequently formed nanoparticles begin to attach to each other in a crystallographically orientated fashion, which is thermodynamically more stable than random attachment. This process results in the development of single crystals (small hexagonal pieces) with an increasingly smooth surface and sharper boundaries. However, the loosely ordered sphere forming in the early stage remains as an aggregation of nanoparticles, not dissolving and thus not re-crystallizing into a more ordered phase thereafter (Fig. 7). The model of vaterite growth is illustrated in Figure 9.

This oriented attachment that occurred after the initial stage of crystallization is a common growth mechanism for nanocrystals [also found in ZnS and $TiO_2$, anatase (Huang et al. 2003; Penn and Banfield 1998)] reducing the overall crystal energy by removing the high-energy surfaces.

Our observations described above reveal a fundamental difference between the growth mode of calcite and vaterite. A plethora of studies convincingly documented the layer-by-layer growth mechanism for calcite crystallization. In particular, in situ imaging techniques such as atomic force microscopy show that the edges of the monolayer calcite remain smooth and maintain strong directionality during growth (Davis et al. 2000; Henriksen et al. 2004; Hillner et al. 1992; Larsen et al. 2010; Teng et al. 1998, 2000; Walters et al. 1997). This suggests that the growth of calcite is, in fact, an ion-by-ion mechanism where the adsorption to an existing kink is highly favored with respect to the generation of a new kink. This implies that growth is highly favored to be spiral growth and the kink (or the continuous filling of the kink that travels around the spiral) leads to a layer-by-layer mechanism. For vaterite growth, however, our data show that there is no such highly-organized molecular attachment scheme, but rather random attachment of nanoparticles with subsequent re-orientation or re-crystallization.

Thermodynamic consideration tells us that the growth of vaterite is likely more entropically favored (in comparison to the growth of calcite) as the clusters at the moment of initial coagulation are randomly oriented. This further suggests that the crystallization of vaterite is probably a kinetically limited process, as a random aggregation of molecular clusters does not significantly lower the free energy of the system, whereas the growth of calcite is controlled more strongly by thermodynamics because oriented and simultaneous attachment of a group of molecules keeps the system's energy gain at the minimum.

The changes of the interior structures of vaterite in different growth stages may explain why the electron-beam resistance of vaterite varies from the early to the final growth stage. Compared to the fast thermal-decomposition of matured crystalline vaterite and calcite under the electron beam (Murooka and Yuan 1993; Xu et al. 1998), our TEM measurements show that nanocrystalline vaterite displays a much higher resistance to ionizing irradiation. Similar phenomena have also been discovered in other materials (Au, $MgGa_2O_4$, TiNi alloy) where particles with smaller particle size (<20 nm) are more irradiation resistant than the larger-grained powders (Zhang et al. 2009). This may be because the aggregations of smaller-sized materials contain more crystallite boundaries, where irradiation-induced defects annihilate more easily. Due to the small size of nanoparticles, defects generated by irradiation within the particles are easy to diffuse to the nanoparticle margins and annihilate there (Chimi et al. 2001; Shen et al. 2007). In contrast, defects in larger particles have a stronger tendency to accumulate in the bulk, leading to radiation-induced decomposition, disorder, and amorphization (Ewing et al. 2004; Lian et al. 2001; Meldrum et al. 1999; Zhang et al. 2009).

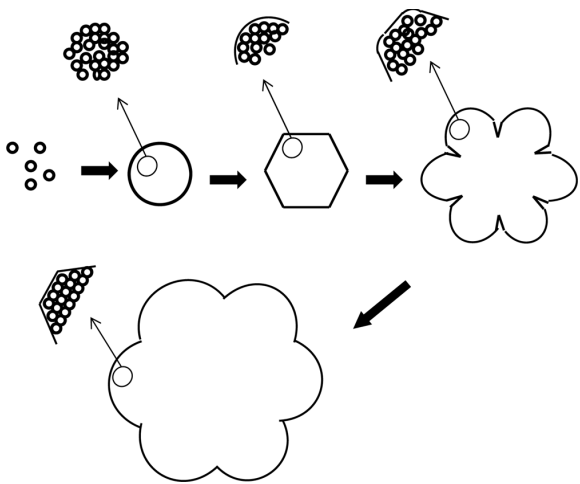

**Figure 10.** Dislocation occurs during the oriented attachments as crystallization proceeds.

## Dislocations caused by the stacking of nanoparticles

Consideration of nanoparticles, instead of ions, as the building blocks may explain why vaterite grains contain a large number of dislocations. While crystal growth from single ions allows for the ordered attachment that leads to the formation of single crystals, the aggregation of larger units such as nanoparticles tends to lead to the formation of grain boundaries that typically do not match epitaxially. Therefore, distortions will occur at the interface to compensate for the mismatch, as also previously reported by Penn and Banfield (1998). This theory may help to explain the common occurrence of dislocations in vaterite among crystal flakes (Fig. 10) and the offset between two overlapping hexagonal platelets (Fig. 5d). The resulting mismatching grain boundaries may also inhibit single crystals growth. The size of the vaterite single crystals (hexagonal platelets), 1–2 μm, is probably the threshold, larger than which the expansion of the continuously regular order of atoms is most likely to be interrupted.

## Growth path of vaterite grains of distinguished morphologies

Another interesting discovery is that the various vaterite shapes at different growth stages in this study are representatives of the distinctive morphologies that have been reported, such as small spheres (Fujiwara et al. 2010; Han et al. 2006a; Watanabe and Akashi 2008), plates (Dupont et al. 1997; Gehrke et al. 2005), and layered flower-like morphologies (Fricke et al. 2006; Heywood et al. 1991). In addition to the similarities of reported shapes to the observed growth morphologies at different growth stages in this study, the respective sizes of crystallites with specific morphologies reported in the literature are also similar to the sizes of the corresponding crystallites of different growth stages reported here. Specifically, most of the spherical balls are about a few micrometers, the plates are on the order of 20 μm, and the flower-like structures are 50–100 μm in diameter.

Another unifying feature of previously reported vaterite (as long as the published images are clear enough to reveal details at the sub-micrometer scale) and the ones shown in this study is that they are composed of nano-sized clusters, independent of the crystal growth method and independent of the resulting vaterite morphology. Vaterite surfaces are all rough and exhibit an appearance of an aggregation of nanoparticles regardless of vaterite formation, whether it was induced by polymers (Colfen and Antonietti 1998; Kasuga et al. 2003; Kim et al. 2005; Kun Park et al. 2004; Shen et al. 2005), amino acids (Grasby 2003; Tong et al. 2004), urea (Wang et al. 1999), bacteria (Rodriguez-Navarro et al. 2007), just from highly supersaturated inorganic solutions (Sohnel and Mullin 1982), or by increasing pH (Han et al. 2006b). Based on the discussion above, we hypothesize that the nanoparticles are the unit building blocks of all types of morphologies of vaterite, and that the growth pathway of vaterite revealed in this study is universal. At conditions where growth limitation develops in the environment, vaterite is not always able to develop into the final stage. For example, decreasing supersaturation, templates, growth inhibitors, and other factors can prevent vaterite from further growth. The fact that the final growth stage is often a flower-like morphology with six "leaves" is likely not to be a result of the hexagonal crystal structure of vaterite because the composing crystallites have more or less random orientation with respect to each other. We consider it more likely that this is the result of diffusion-limited growth as in the more macroscopic growth of snowflakes.